\documentclass[preprint,epsf,aps]{revtex4}
\usepackage{graphicx,graphics,color,epsfig}
\begin{document}
\title{Fault tolerant quantum key distribution protocol 
with collective random unitary noise 
}
\author{Xiang-Bin Wang\thanks{email: wang$@$qci.jst.go.jp} 
\\
        Imai Quantum Computation and Information project, ERATO, JST\\
Daini Hongo White Bldg. 201, 5-28-3, Hongo, Bunkyo, Tokyo 113-0033, Japan}
\begin{abstract}
\thispagestyle{empty}
We propose an easy implementable prepare-and-measure protocol for robust quantum 
key distribution with photon polarization. 
The protocol is fault tolerant against
collective random unitary channel noise. 
The protocol does not need any collective quantum measurement or quantum memory.
A security proof and a specific linear optical realization 
using spontaneous parametric down conversion are given. 
\end{abstract}
\maketitle 
{\it Introduction.}
Quantum key distribution (QKD)\cite{BB,gisin} is one of the most important 
application
of the subject of quantum information. In constrast to classical 
cryptography, the security of QKD is guaranteed by elementary principles
of quantum mechanics, and therefore the unconditional security can be achieved.
For security, we have to distill out a shorter final key, since
Eavesdropper (Eve) may pretend her disturbance to be the noise
from the physical channel. If the noise is too large, no final key can be 
obtained. To overcome this, one needs to design new fault tolerant 
protocols or new physical realizations for quantum key distribution. There are two
approaches two this problem: one is to find a new protocol which raises the
threshold of channel noise unconditionally, such as the protocol with 2 way
classical communications\cite{gl,chau,wang}; the other way is first to study
the noise pattern and then find a way to remove or decrease the noise 
itself, such as the various method to cancel the collective 
errors\cite{ekert,walton,boil}.
So far there are 
various realizations using either the phase-coding\cite{gisin,gisin1}
or the polarization information of single photons\cite{but,zeilinger}.
Those protocols using the phase-coding requires collective measurement at Bob's
side.
There are also proposals to remove the collective random unitary noise from
the channel\cite{ekert,walton,boil}.
 
Here we raise a new proposal to reduce the channel errors, or, equivalently, to
raise the noise threshold. Our method does not require Bob to take any collective
measurement. 
Our method is based on the widely accepted assumptions that
the flipping errors of polarization (mainly) come from the random
rotation by the fiber or
the molecules in the air, with the degree of the rotation fluctuating 
randomly. Also, if several qubits are transmitted simultaneously and they
are spatially close to each other, the random unitaries to each of them
must be identical, i.e., the error of the physical channel is {\it collective}.
\\{\it Main Idea.}
Consider an arbitrary collective random unitary $U$ which satisfies
\begin{eqnarray} \nonumber
U|0\rangle= \cos\theta |0\rangle + e^{i\phi} \sin\theta |1\rangle;
\\U|1\rangle= e^{i\Delta}(-e^{-i\phi}\sin\theta |0\rangle +
\cos\theta |1\rangle
\label{dephase}
\end{eqnarray}
 Here $|0\rangle,|1\rangle$ represent for horizontal
and vertical polarization states respectively. 
Note that the parameters $\Delta,\phi$ and $\theta$ fluctuate with time, 
therefore one has no way to make unitary compensation to a single qubit.
However, the channel unitary error is a type of collective error to all
qubits sent simultaneously, therefore it is possible to 
send qubits robustly because the collective errors on different qubits 
may cancel each other. 
With such type of collective unitary errors, we shall take the QKD in the subspace of two-qubit state of 
\begin{eqnarray}
S=\{|01\rangle, |10\rangle\}.
\end{eqnarray} 
In particular, we let
Alice prepare and send Bob two-qubit states randomly chosen from
$|01\rangle,|10\rangle,|\psi^\pm\rangle=\frac{1}{\sqrt 2}(|01\rangle \pm |10\rangle)$. Although state $|\psi^-\rangle$ keeps unchanged under the collective
unitary errors\cite{kwiat}, the other 3 states do not keep unchanged. However, in our 
protocol, we shall let Bob first take a parity check to the two-qubit state to
see whether it belongs to subspace   $S$. If it does, he accepts it, if it does not, he discard it. The key point here is that,
although  the 2-qubit states could be distorted by the collective random 
unitary, most often
the distortion will drive the codes out of subspace $S$ therefore the distorted codes will be discarded by the protocol itself. 
The error rates to those {\it accepted} codes are normally 
small, provided that the channel noise are mainly from the collective 
unitary and the averaged value $\theta$ is not too large. 
For example, our protocol gives a good key rate if the averaged value 
$|\sin\theta|$ is 1/2.  
(The dispersion, $\phi$ value can be arbitrarily large.)     
Explicitly, any collective rotation cannot exchange states $|\psi^+\rangle$ and $|\psi^-\rangle$, it can only drive $|\psi^+\rangle$ out of the subspace $S$. However, any state outside of $S$ will be rejected, as required by our protocol. Therefore the rate of flipping between 
$|\psi^+\rangle$ and $\psi^-\rangle$ (phase-flip rate) 
is zero. A collective rotation $U$ will also take the following effects:
\begin{eqnarray}\nonumber
U^{\otimes 2} |01\rangle=
U|0\rangle \otimes U|1\rangle\\
=\cos^2\theta |01\rangle - \sin\theta\cos\theta
(e^{-i\phi}|00\rangle + e^{i\phi}|11\rangle) +
\sin^2\theta |10\rangle .
\end{eqnarray}
\begin{eqnarray}\nonumber
U^{\otimes 2} |10\rangle=
U|1\rangle \otimes U|0\rangle\\
=\cos^2\theta |10\rangle - \sin\theta\cos\theta(
e^{-i\phi}|00\rangle + e^{i\phi}|11\rangle) +
\sin^2\theta |01\rangle .
\end{eqnarray} 
Since the states outside the subspace $S$ will be discarded, the
net flipping between rate between $|01\rangle$ and $|10\rangle$ 
(bit-flip rate) 
$r_b= \frac{\sin^4\theta}{\cos^4\theta +\sin^4\theta}$. Therefore, if the average rotating angle is small, the flipping rate $r_b$ will be also small.
(If we directly use BB84 protocol, the bit-flip rate is $\sin^2\theta$, one magnitude order larger than ours.)
Moreover, in the ideal case that all flips come from the random rotation,
since the phase-flip rate is zero, one can $always $ distill some bits of final key provided that $r_b\not= 1/2$. 
The key rate is 
$1+r_b\log_2 r_b + (1-r_b)\log_2 (1-r_b)$. 
Note that if $r_b>1/2$ one can simply reverse all bit values given by
$|01\rangle$ and $|10\rangle$ and also distill some bits of final key.
In practice, if $\theta$ does not change too fast, we can divide the data into many blocks, say, each block
contains the data with several seconds. We inverse the bit-values of those blocks with larger than 1/2 error rate
after the decoding. Note that we assume the phase-flip error to be always very small by our protocol.

Boileau et al\cite{boil} has proposed a protocol
with the collective random unitary error model recently. Our work differs from ref.\cite{boil}
in the following aspects:
1). The main idea is different. Ref.\cite{boil} uses the fact that state 
$|\psi^-\rangle=\frac{1}{\sqrt 2}(|01\rangle - |10\rangle)$ is invariant under
whatever rotations therefore the linear combinations of a few $|\psi^-\rangle$ at different positions will work robustly. We use a subspace of two qubit state. Our states are
not always invariant under random rotations, however, the randomly rotation can
drive the original state out of the specific subspace and never or rarely
switch any two states inside the subspace. After Bob 
discards all those transmitted
codes outside subspace $S$, the phase-flip error will be totally removed 
and the bit flip error will be significantly decreased. 
2). The method is different. The protocol given by Boileau et al requires
3-qubit or 4-qubit entangled states, which could be technically difficult by
currently existing technology. Our protocol only requires  2-qubit states
which can be produced effectively.
3). The result is different.
Since our protocol is BB84-like\cite{BB},
we don't have to worry about the channel loss in practice. Boileau's
 protocol is likely to be
 undermined by the
channel loss, since it is B92-like\cite{B92,tamaki}. 
In practice, the 
lossy rate for their protocol could be very high. Since they use at least
3 qubits to encode one, the joint survival rate is very low.
\\{\it Protocol 1 and Security Proof.} For clarity, we now give a protocol with collective measurements first and then
reduce it to a practically feasible protocol without any collective 
measurements.
\\{\it Protocol 1}
{\bf 1: Preparation of the encoded BB84 states.} 
Alice creates a number of  single qubit states, each of them is
randomly chosen from 
$\{|0\rangle,|1\rangle,|\pm\rangle\}$. She put down each one's preparation basis and bit value:
state $|0\rangle,|+\rangle$ for bit value 0, the other 2 states are for 1. 
She also prepares  
 ancillas which are all in state $|0\rangle$.
She then encodes each individual qubit with an ancilla 
into a 2-qubit code through the following CNOT operation:
$|00\rangle\longrightarrow |01\rangle; |10\rangle\longrightarrow |10\rangle;
 |11\rangle\longrightarrow |11\rangle;  |01\rangle\longrightarrow |00\rangle$.
The second digit in each  state is for the
ancilla. Such encoding operation changes $(|0\rangle,|1\rangle)$ into $(|01\rangle,|10\rangle)$ and
$|\pm\rangle$ into $|\psi^{\pm}\rangle$.
{\bf 2: State transmission.} Alice sends those 2-qubit codes to Bob.
{\bf 3: Error-rejection and Decoding.}  Bob takes the same CNOT operation
as used by Alice in encoding. He then measures the second qubit in $Z$ basis:
if it is $|1\rangle$, he discards both qubits  and notifies Alice; 
if he obtains $|0\rangle$,
he measures the first qubit in either $X$ basis or $Z$ basis
and records the basis as his ``measurement basis''
in the QKD protocol. 
The bit-value of a code is determined by the measurement outcome of the first
qubit after decoding,  $|0\rangle,|+\rangle$ for bit value
0, $|1\rangle,|-\rangle$ correspond to 1. 
{\bf 4: Basis announcement.} 
Through public discussion, they discard all those decoded qubits 
with different measurement bases in two sides.
{\bf 5: Error test.} They announce the values of some randomly chosen $X$ bits and the same 
number
of $Z$ bits. If too many values disagree, they abort the 
protocol. Otherwise they distill the remained  ``$Z$ bits'' for the final key.     
{\bf 6: Final key distillation.} Alice and Bob distill the final key from the remained 
``$Z$ bits'' by 
 the classical
CSS code\cite{shorpre}.\\
 The unconditional security here is equivalent to that
of BB84\cite{BB,shorpre} with a lossy noisy channel: {\it Protocol 1} 
can be regarded as an encoded BB84 protocol
 with additional steps of encoding, error rejection and decoding. 
If  Eve. can attack {\it Protocol 1} successfully 
with operation $\hat A$ during
the stage of codes transmission, she can
also attack BB84 protocol successfully with 
\begin{eqnarray}
\hat A'=\hat E \longrightarrow\hat A\longrightarrow \hat R \longrightarrow\hat D
\end{eqnarray}
during the qubit transmission and then pass the decoded qubit to Bob, where 
$\hat E,\hat R,\hat D$ are encoding, error rejection and quantum decoding, respectively.
(The operation of encoding, error rejection or decoding does not requires any information about the unknown
state itself.) 
Obviously, BB84 protocol with attack $\hat A'$ is identical to {\it Protocol 1} with attack $\hat A$. 
To Alice and Bob, BB84  protocol with Eve's attack $\hat A'$ is just a BB84 protocol with   
a lossy channel.  (Eve must discard some codes in the error rejection step.) 
Therefore {\it Protocol 1} must be secure, since 
 BB84 protocol is unconditional secure even with a lossy channel. 
\\{\it  Protocol 2.} Though we have demonstrated the unconditional security of {\it Protocol 1}, we do not
directly use {\it Protocol 1} in practice since it requires the local CNOT 
operation
in encoding and decoding. We now reduce it to another protocol without any collective operations. First, since there are only 4 candidates in the set of
BB84 states, instead of encoding from BB84 states, Alice may directly produce 4 random states of $|01\rangle,|10\rangle, |\psi^+\rangle,|\psi^-\rangle$. 
Note that except for Alice herself, no one else can see whether the 
two-qubit codes in transmission are directly produced or 
the encoding result from BB84 states.
One may simply  produce the states of those 2-qubit codes by the spontaneous
parametric down conversion\cite{spdc,para}. Second, in the decoding
and error rejection step,
Bob can carry out the task by {\it post-selection}.
For all those codes originally in state $|01\rangle$ or $|10\rangle$, Bob can simply
take local measurements in $Z$ basis to each qubits and then discard those
outcome of $|0\rangle\otimes |0\rangle$ or $|1\rangle\otimes |1\rangle$ and
only accepts the outcome $|0\rangle\otimes |1\rangle$ which is regarded as
a bit value 0 and $|1\rangle\otimes |0\rangle$ which is regarded as bit 
value 1. The net flipping rate between $|01\rangle$ and $|10\rangle$ 
is regarded
as bit-flip rate. 
The non-trivial point is the phase-flip rate, i.e., the net flipping
rate between states $|\psi^{\pm}\rangle$.
Note that all these codes only take the role of indicating the phase-flip
rate, we don't have to know explicitly which one is flipped and which one is not flipped.
Instead, we only need to know the average flipping rate between $|\psi^\pm\rangle$.
To obtain such information, we actually 
don't have to really carry out the error rejection and 
decoding steps to each of these codes. What we need to do is simply to answer
what the flipping rate $would$ be if Bob  really $took$ the 
error rejection step and decoding step to each codes of $|\psi^{\pm}\rangle$.
One straight forward way is to let Bob 
take a Bell measurement to each code which were
in state $|\psi^\pm\rangle$ originally.({\it We shall call them $\psi^+$ codes or
$\psi^-$ codes hereafter.}) For example, consider $\psi^-$ codes, after transmission, 
if the distribution over 4 Bell 
states $|\psi^+\rangle, |\psi^-\rangle, |\phi^+\rangle, |\phi^-\rangle$
 are $p_{\psi^-}, p_{\psi^+}, p_{\phi^+},p_{\phi^-}$, respectively after the Bell measurements,  we conclude that the channel flipping rate of $|\psi^+\rangle\longrightarrow |\psi^-\rangle$ is $p_{\psi^-}/(p_{\psi^+}+p_{\psi^-})$. This rate is equivalent to the flipping rate of
$|+\rangle\longrightarrow |-\rangle$ in BB84 protocol. Note that the rate of
$q_{\phi^\pm}$ have been excluded here since their corresponding states are outside of the subspace $S$ and should 
be discarded by our protocol. 
 
Bell measurement is not the unique way to see the distribution over 4 Bell states for a set of states. We can also simply divide the set into 3 subsets and
take collective measurements $ZZ$ to subset 1, $XX$ to subset 2, and
$YY$ to subset 3. We can  then $deduce$ the distribution over the 4 Bell states. Here  $ZZ,XX,YY$ are parity measurements
to a two-qubit code in $Z,X,Y$ basis, respectively. ($Y:$ measurement basis of
$\{|y\pm\rangle = \frac{1}{\sqrt 2}(|0\rangle\pm i|1\rangle)\}$.)
Note that here classical
statistics works perfectly because all these collective measurements commute
\cite{lc,wang0}.
 These collective 
measurements can be simply replaced by local measurements to each qubits since
once we have done the results of local measurements of 
$Z\otimes Z,X\otimes X, Y\otimes Y$ we know the parity information. (In this paper,  $Z\otimes Z$ represents
a local measurement to each qubit in $Z$ basis; $ZZ$ represents a collective measurement for the parity in $Z$
basis.)

Before 
going into the reduced protocol, we show the explicit relationship between
the phase-flip rate and the local measurement results.
Note that Bob has randomly divided all the received 2-qubit codes into 3
subsets and he will take local measurement $Z\otimes Z,X\otimes X, Y\otimes Y$
to each of the qubits of each codes in subset 1,2,3, respectively.
Consider all $\psi^-$ codes first. Denote  
$\epsilon_z,\epsilon_x,\epsilon_y$ for
the rate of wrong outcome for $\psi^-$ codes in subset 1,2,3, respectively, i.e. the rate
of codes whose two qubit has the same bit values in basis $Z,X,Y$, respectively.
Given values $\epsilon_{z,x,y}$ we immediately have
\begin{eqnarray}
p_{\phi^+}+p_{\phi^-}= \epsilon_z\\
p_{\psi^+}+p_{\phi^+}= \epsilon_x\\
p_{\psi^+}+p_{\phi^-}=\epsilon_y.
\end{eqnarray}  
Our aim is only to see the flipping rate from $|\psi^-\rangle$ to 
$|\psi^+\rangle$, other types of errors are discarded since they have gone out of the given subspace $S$.
The net flipping rate from $|\psi^-\rangle$ to 
$|\psi^+\rangle$ is
\begin{eqnarray}\label{test1}
t_{\psi^-\rightarrow \psi^+}=\frac{p_{\psi^+}}{p_{\psi^-}+p_{\psi^+}}
=\frac{\epsilon_x+\epsilon_y-\epsilon_z}{2(1-\epsilon_z)}.
\end{eqnarray}
In a similar way we can also have the formular for the value
of $t_{\psi^+\rightarrow \psi^-}$, the flipping 
rate from $|\psi^+\rangle$ to $|\psi^-\rangle$:
\begin{eqnarray}
t_{\psi^+\rightarrow \psi^-}
=\frac{{\epsilon'}_x+{\epsilon'}_y-{\epsilon'}_z}{2(1-{\epsilon'}_z)}.
\end{eqnarray}
Here ${\epsilon'}_{x,y,z}$ are rate of wrong outcome in local measurement basis
$X\otimes X, Y\otimes Y, Z\otimes Z$, respectively, to all codes originally
in $|\psi^+\rangle$.
The total phase-flip error is
\begin{eqnarray}\label{phase}
t_p=\frac{t_{\psi^-\rightarrow \psi^+}+t_{\psi^+\rightarrow \psi^-}}{2}.
\end{eqnarray}  
{\it Protocol 1} is now replaced by the following practically feasible protocol without any
collective measurement:
 \\{\it Protocol 2}
{\bf 1: Preparation of the encoded BB84 states.} 
Alice creates a number of 2-qubit states and each of them are randomly chosen from
$\{|01\rangle,|10\rangle, |\psi^{\pm}\rangle\}$. 
 For each 2-qubit code, she puts down 
``Z basis'' if
it is in state $|01\rangle$ or $|10\rangle$ or
 ``X basis'' ($\{|\pm\rangle\}$) if it is in one of the states 
$\{\frac{1}{\sqrt 2}(|01\rangle\pm |10\rangle)$. 
For those code states of  $|01\rangle$ or  
$\{\frac{1}{\sqrt 2}(|01\rangle+|10\rangle)$, she denotes a bit value 0 ;
for those code states of $|10\rangle$ or  
$\{\frac{1}{\sqrt 2}(|01\rangle-|10\rangle)$, she denotes a bit value 1.
{\bf 2: Transmission.} Alice sends all the 2-qubit codes to Bob.
{\bf 3: Measurement.} To each code, Bob measures the two qubits in a basis
 randomly chosen from
$\{Z\otimes Z, X\otimes X, Y\otimes Y\}$. For example, if he happens to choose basis $Z\otimes Z$ for a certain code, 
he measures each qubit of that code in $Z$ basis. 
{\bf 4: Rejection of wrong results.} Alice announces her ``preparation basis'' for each codes.
Bob announces his measurement basis to  each codes. For those
codes originally prepared in $|01\rangle$ or $|10\rangle$, they discard the
results if Bob has used a basis other than $Z\otimes Z$. 
They also discard all codes outside the subspace $S$.
{\bf 5: Error test.} To all the survived results, 
they announce some bit values of codes originally in $|\psi^+\rangle$ or $|\psi^-\rangle$. 
From the announced results they can calculate the phase-flip rate by formula(\ref{phase}).
They can also estimate the bit-flip rate by annoucing 
some results of those survived codes which are originally in $|01\rangle$ or $|10\rangle$. 
{\bf 6: Final key distillation.} Alice and Bob distill the final key from the remianed ``$Z$ bits'' 
 by using the classical
CSS code\cite{shorpre}.  (Since they only use ``$Z$ bits'' for final key distillation, Alice can choose 
``$Z$ basis'' more frequently than ``$X$ basis'' in Step 1.)
\\{\it Physical Realization of Protocol 2.}
 There are
two parts in the realization. One is the source for the required 4 different
2-qubit states at Alice's side. The other is the measurement device at Bob's
side. Both of them can be realized with simple linear optical devices. 
The requested source states can be generated by 
SPDC process\cite{spdc,para} as shown in figure 1. The measurement with random basis
at Bob's side
can be done by a polarizing beam splitter(PBS) and a rotator driven electrically, as shown in figure 2.
\begin{figure}
\epsffile{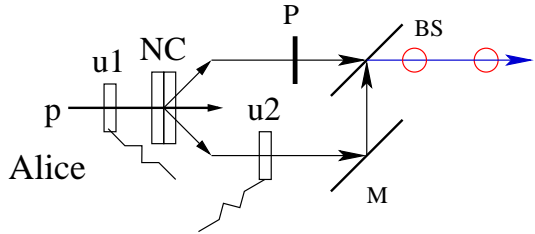}
\vskip 5pt
\caption{(Color online)  The source of two-qubit state.
 P: $\pi/2$ rotater. BS: beam splitter, M: mirror, NC: nonlinear crystal,
 p: pump light in horizontal polarization, u1: unitary rotator, 
u2: phase shifter. u1 takes the value of 0, $\pi/2$, $\pi/4$ to produce 
state
$|01\rangle, |10\rangle, |\psi^+\rangle$, respectively. u2 can be either
$I$ or $\sigma_z$. }\label{source}
\end{figure}  
\begin{figure}
\epsffile{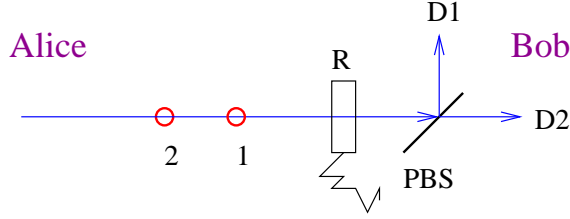}
\vskip 5pt
\caption{(Color online) Measurement device at Bob's side.
The rotator R offers a random
rotation to both qubits in the same code. Each time, rotation is
randomly chosen from unity, $(|0\rangle,|1\rangle)\longrightarrow (|+\rangle,|-\rangle)$, 
$(|0\rangle,|1\rangle)\longrightarrow (|y+\rangle,|y-\rangle)$.
The event of two clicks on one detector (D1 or D2) show that the 2 qubits of the code have the same bit value,
 two clicks on different detectors show that the
2 qubits have different bit values.
\label{scheme2} }
\end{figure}
\\{\it Another protocol for robust QKD with swinging objects.}
In some cases, especially in free space, the dispersion can be small 
while the random
rotation angle $\theta$ can be large. We consider the extreme case
that $\phi$ in unitary $U$ is 0, or otherwise can be compensated to almost
0, but $\theta$ is random and can be arbitrarily large. The swinging angle
of an airplane can be very large in certain case. We can exactly use the 
collective unitary 
model, with all elements in $U$ being real if there is no dispersion.  
Then we have a better method. It is well known that both states $|\phi^+\rangle$ and $|\psi^-\rangle$ are invariant under whatever real rotation.
Any linear superposed state of these two are also invariant. Therefore we use
the following for states $\{|\bar 0\rangle=|\phi^+\rangle,|\bar 1\rangle=|\psi^-\rangle; |+'\rangle=\frac{1}{\sqrt 2}(|\bar 0\rangle + |\bar 1\rangle)=
\frac{1}{\sqrt 2}
(|0\rangle|+\rangle -|1\rangle|-\rangle); |-'\rangle=\frac{1}{\sqrt 2}
(|0\rangle|-\rangle +|1\rangle|+\rangle)\}.$ 
Bob need not take any collective measurement to determine the bit value.
If he chooses ``Z'' basis, he measure each of the two qubits in $Z$ basis,
$00$ or $11$ for bit value 0 while 01 or 10 for bit value 1. If he chooses
``X'' basis, he measures the first qubit in $Z$ basis and the second in
$X$ basis, $|0\rangle|+\rangle$ or $|1\rangle|-\rangle$ for bit value 0 and
 $|0\rangle|-\rangle$ or $|1\rangle|+\rangle$ for bit value 1. There is no 
error-rejection step here because it is expected to be no error 
after decoding, given the real rorarion channel. Even for
the QKD with fixed object there is still a little bit advantage: they do
not need take any bases allignment with each other. Each of them only need to
make sure their local measurement bases are BB84-like, i.e., 
the inner product of two bases are $\frac{1}{\sqrt 2}$. 
\\  
{\it Concluding remark.} We have given a robust QKD protocol in polarization
space given that the collective random unitaries are dominant channels errors.
 Our protocol can obviously be extended to the 
6-state-like protocol\cite{bruss} if we add one more candidate state 
of $\frac{1}{\sqrt 2}(|0\rangle\pm i|1\rangle)$in the source.\\  
{\bf Acknowledgement:} I thank Prof. H. Imai for support. I thank  J.W. Pan,
B.S. Shi and A. Tomita
for discussions.
\\{\em Note Added:} 
After the work was completed, a 
different novel protocol\cite{b118} for robust QKD has drawn our attention.


\begin{thebibliography}{99}
\bibitem{BB}
C. H. Bennett and G. Brassard, 
{\em Proceedings of IEEE International Conference on Computers, 
Systems and Signal Processing, Bangalore, India, 1984},  (IEEE Press,
1984), pp. 175--179;
C.H. Bennett and G. Brassard,
IBM Technical Disclosure Bulletin {\bf 28}, 3153--3163 (1985).
\bibitem{gisin} N. Gisin, G. Ribordy, W. Tittel, and H. Zbinden,
 Reviews of Modern Physics, vol. 74, pp. 145-195.
\bibitem{gl} D. Gottesman and H.-K. Lo, IEEE Transactions on
 Information Theory, 49, 457(2003).
 \bibitem{chau} H. F. Chau, Phys. Rev. A66, 060302(R) (2002).
\bibitem{wang} X. B. Wang, Phys. Rev. Lett., 92, 077902(2004).
\bibitem{ekert}G.M. Palma, K.A. Suominen, and A.K. Ekert, Proc. R. Soc. London A 452, 567(1996).
\bibitem{walton} Z. D. Walton et al,  Phys. Rev. Lett., 91, 087901(2003).
\bibitem{boil} J.C. Boileau, D. Gottesman, R. Laflamme, D. Poulin and R.W. 
Spekkens, Phys. Rev. Lett. 92, 17901(2004).
\bibitem{B92}C. H. Bennett, Phys. Rev. Lett. 68, 3121(1992).
\bibitem{tamaki} K. Tamaki, M. Koashi, and N. Imoto, Phys. Rev. Lett. 90, 167904(2003).
\bibitem{kwiat} P.G. Kwiat et al, Science 290, 498(2000).
\bibitem{shorpre} P. W. Shor and J. Preskill, Phys. Rev. Lett., vol. 85,441(2000).
\bibitem{spdc}P.G. Kwiat et al, Phys. Rev. A60, R773(1999).
\bibitem{para} P. G. Kwiat, K. Mattle, H. Weinfurter, A. Zeilinger, A.V. Sergienko, and Y. H. Shih, Phys. Rev. Lett. 75, 4337(1995).
\bibitem{lc} H.K. Lo and H.F. Chau, Science, 283, 2050(1999).
\bibitem{wang0}X.B. Wang, quant-ph/0403058.
\bibitem{bruss} D. Bruss, Phys. Rev. Lett. 81, 3018(1998).
\bibitem{gisin1}D. Stucki et al, New J. Phys., 4, 41(2002).
\bibitem{but}W. T. Buttler et al, Phys. Rev. Lett. 81, 3283(1998)
\bibitem{zeilinger} M. Aspelmeyer et al, Science, 301, 621(2003); M. Aspelmeyer et al, IEEE J of Selected Topics in Quant. Electronics, 9, 1541(2003); 
G. J. Rarity et al, New J. Phys. 4, 82(2002) 
\bibitem{b118}J.-C. Boileau, R. Laflamme, M. Laforest, C. R. Myers,
quant-ph/0406118, published in Phys. Rev. Lett., 93, 220501, Nov 26, 2004.
\end{thebibliography}
\end{document}